\documentclass[9pt]{elife}

\usepackage{lipsum} 
\usepackage[version=4]{mhchem}
\usepackage{siunitx}
\DeclareSIUnit\Molar{M}

\title{Mechanistic insights into the spatial organization of RNA polymerase proteins and the chromosome in {\em E.\ coli} cells.}

\author[1*]{Debarshi Mitra}
\author[1,2,3*]{Jens-Uwe Sommer}
\affil[1]{Leibniz-Institut für Polymerforschung Dresden, Hohe Strasse 6, 01069 Dresden, Germany}
\affil[2]{Institute for Theoretical Physics, TU Dresden, 01062 Dresden, Germany}
\affil[3]{Cluster of Excellence Physics of Life, TU Dresden, 01307 Dresden, Germany}

\corr{mitra@ipfdd.de}{DM}
\corr{sommer@ipfdd.de}{JUS}

\begin{document}

\maketitle

\begin{abstract}
Along the bacterial chromosome, regions called {\em rrn operons} contain genes that are transcribed into ribosomal RNA. These operons are among the most transcriptionally active sites in the genome. It has been observed in {\em E.\ coli} that RNA polymerase (RNAP), while binding to these loci along the chromosome during transcription, forms clusters, leading to spatial colocalization of the operons within the cell. Recent experimental evidence suggests that liquid–liquid phase separation contributes to the formation of RNAP clusters, with the antitermination factor NusA playing a key role. 
We present a simulation model to investigate the mechanisms underlying the formation of these biomolecular condensates. We propose that mutual attraction between NusA proteins, which exhibit a miscibility gap at higher concentrations, drives condensate formation {\em via} a polymer-assisted condensation pathway, and we demonstrate how these condensates promote the colocalization of {\em rrn operons}. {Our results reconcile disparate experimental observations on the organization of Nus proteins and the chromosome.}
\end{abstract}

\section{Introduction}
 In the last decade, biomolecular condensates have been at the forefront of biophysical research \citep{Hyman2014,Shin2017}. These condensates act as membraneless organelles inside cells and form {\em via} liquid–liquid phase separation (LLPS) \citep{Hyman2014,roadmap}. Biomolecular condensates are known to mediate a wide range of biological processes \citep{Brangwynne2009,Feric2016,Strom2017,Mukherjee2026} and a variety of diseases have been linked to their aberrant formation or regulation \citep{Ripin2023,Alberti2019,Gao2023}.

In cells of higher order organisms, it is well established that the nucleolus is a membraneless organelle formed {\em via} LLPS \citep{nucleolus,nucleolus2,nucleolus3} that acts as a site of ribosome synthesis.  The nucleolus contains ribosomal RNA genes, nascent rRNA transcripts, RNA polymerase (hereafter referred to as RNAP), and numerous auxiliary proteins required for ribosome biogenesis \citep{Harmon2020}. By contrast, the role of biomolecular condensates in bacterial cells, and particularly their consequences for the organization of chromosomes, has remained largely elusive \citep{Hodgins2025}.

The organization of chromosomes in bacterial cells continues to be a long-standing research problem. In the case of {\em E.\ coli}, the chromosome has a length of approximately $\sim 1\, \text{mm}$ when fully stretched, yet it is confined within a cell of length $\sim 2\, \mathrm{\mu m}$. To achieve this high level of compaction, the {\em E.\ coli} chromosome is organized and compacted at multiple length scales. Furthermore, the chromosome participates in replication and transcription, which rely on specific interactions between various proteins and chromosomal loci. 

Two complementary classes of experimental techniques: Hi-C \citep{Lioy2018} and fluorescence-labeling-based methods such as FISH (Fluorescence In Situ Hybridization) \citep{Cass2016,caul_loci,Wiggins2018,Youngren2014}, have been widely used to probe the spatial organization of bacterial chromosomes. Hi-C experiments provide information about the ensemble-averaged likelihood that two genomic segments are in spatial proximity, where the effective spatial range of proximity detection is determined by the experimental protocol, particularly the crosslinking and ligation steps. FISH experiments on the other hand, provide information about the spatial positioning and dynamics of specific tagged genomic loci. Together, these two complementary approaches provide comprehensive insights into the static and dynamical aspects of chromosome organization.

\begin{figure*}

\hskip2.7cm
\includegraphics[width=0.55\columnwidth,angle=0]{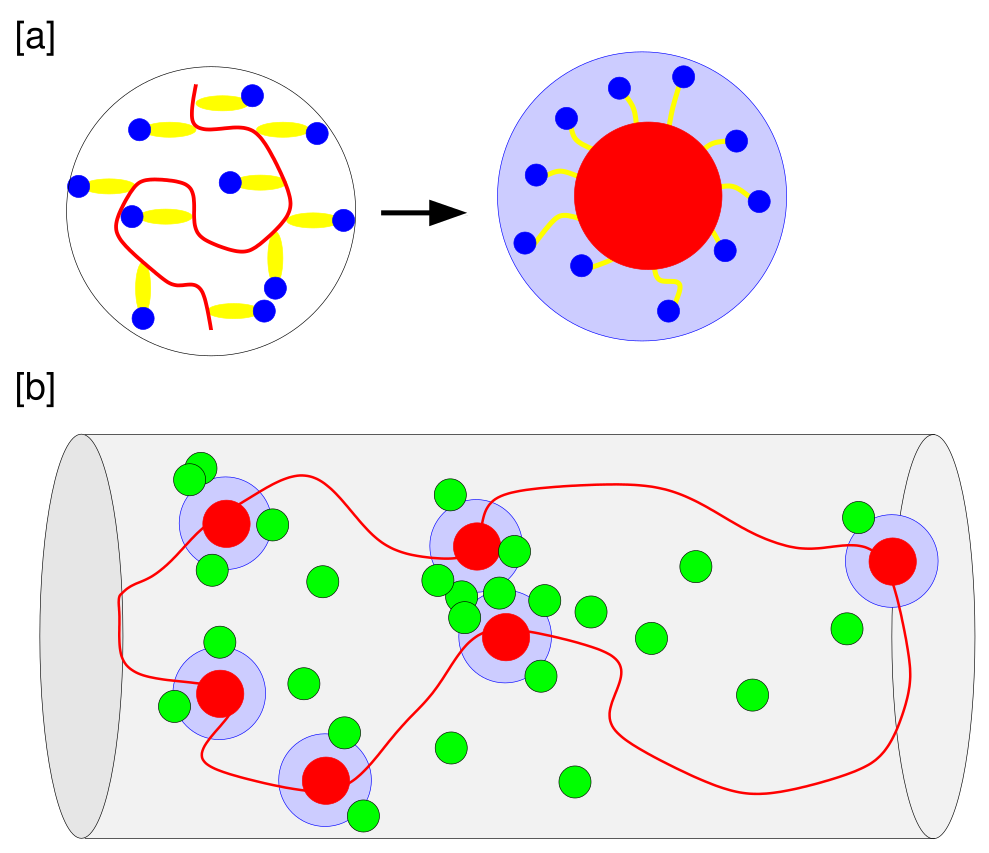}
\caption{\label{fig1}
{\bf Schematic illustration of the coarse-graining and our model:} (a) Left: the rrn operon sequence (red strand) binds to RNAP (yellow) and `bound' Nus proteins (blue); 10 of 42 (both Nus and RNAP) proteins are shown. Right: the operon is mapped onto an effective monomer in our model. RNAP acts as a linker connecting bound Nus proteins to the operon. The Nus proteins form a corona around the operon. (b) Sketch of the coarse-grained model of the chromosome inside the cylindrical {\em E.\ coli} cell. The free Nus proteins are shown in green display attractive interaction and are also attracted by the corona particles attached to the {\em rrn operons}. }
\end{figure*}


The factors that drive genome organization are closely related to the cellular processes such as replication and transcription. Transcription is carried out by RNAPs, which binds to specific genetic loci along the chromosome. 
Along the chromosome contour, there exist specific loci known as {\em rrn operons}, which contain genes encoding ribosomal RNA (rRNA). These {\em rrn operons} (all except one) have been observed to spatially colocalize in {\em E.\ coli} cells \citep{Gaal}, largely independent of growth conditions. However, this colocalization is statistical in nature: if one considers any pair of {\em rrn operons}, the average spatial distance between them is smaller than expected for two genomic loci separated by a similar genomic distance along the chromosome \citep{Gaal}. However, contact measurements obtained from Hi-C experiments indicate that there are no stable or long-lived contacts between distant genomic segments along the {\em E.\ coli} chromosomal contour \citep{Lioy2018}. { This discrepancy may arise due to a number of reasons: The Hi-C method employs formaldehyde to cross-link proximal chromosomal sites \citep{hic_original}. The use of formaldehyde may disrupt or dissolve the operon cluster. Furthermore, the distance between the {\em rrn operons} may still be sufficiently large such that their spatial proximity is not captured by Hi-C, which relies on cross-linking of chromosomal segments that are in close spatial proximity.  }
{ It should also however be noted that several other FISH based experiments do not report the colocalization of {\em rrn operons} \citep{Wiggins2010,Cass2016}.  Recent evidence obtained by a different experimental protocol called the {\em Mu} method, however once again affirmed the colocalization of {\em rrn operons} \citep{mu1,Ho2025}.}


It has recently been established that RNAPs form distinct clusters (condensates) inside {\em E.\ coli} cells \citep{rnap_llps,achilles}. This is observed even for {\em E.\ coli} strains where five of the seven {\em rrn operons} have been artificially deleted \citep{achilles}. The work of \cite{rnap_llps} further suggests that phase separation is driven by an anti-termination factor protein called NusA (hereafter called Nus).  Nus binds to RNAP (which is bound to DNA) during transcription and helps in the efficient transcription of the ribosomal RNA genes. The authors of \citep{rnap_llps} further showed that the Nus phase separates {\em in-vitro} in the presence of a crowding agent, albeit at much higher concentrations than what is seen {\em in-vivo}.To visualize how the RNAPs and Nus are arranged in the cell during transcription, refer to the model schematic in Fig. \ref{fig1}.  { In particular, the work of \cite{rnap_llps} established that the population of Nus proteins {\em in vivo} exhibit three
distinct dynamical states: a freely diffusing population, a RNAP-bound population and a third population with intermediate liquid-like mobility that is consistent with the presence of a condensate.  }

The work of \cite{xiao19} established a key link between the clustering of RNAPs and transcription. They showed that the clustering of RNAPs diminish significantly when transcription is inhibited \citep{xiao19}. During transcription, a large fraction ( $\approx 30\%$) of RNAPs bind exclusively to the sites of {\em rrn operons}, which are among the most transcriptionally active sites in the chromosome \citep{Bremer2008}. This indicates that the binding of RNAPs to genetic sites (and especially to the sites of {\em rrn operons}) plays a key role in inducing the clustering of Nus and RNAPs in the cell. 

As mentioned before, it was shown in \cite{rnap_llps} that Nus only phase separates {\em in-vitro} at concentrations much higher than what is present {\em in vivo}. It is still worth asking if phase separation can occur {\em in vivo} due to the  binding of Nus to RNAPs, which in turn are bound to genetic sites during transcription. 
In this context, previous work \citep{Sommer_PAC} has demonstrated that proteins that interact weakly with a long flexible polymer, such as a chromatin, can undergo phase separation even when the bulk protein solution lies outside its miscibility gap. This mechanism, termed Polymer Assisted Condensation (PAC), arises because weak attractive interactions between proteins and the polymer lead to an enrichment of proteins in the vicinity of the polymer. As a result, the local protein concentration near the polymer can become significantly higher than in the surrounding bulk solution. When this local concentration exceeds the threshold for phase separation, a condensate nucleates on the polymer. Thus, the polymer effectively acts as a scaffold that promotes condensation. Motivated by this mechanism, we propose that Nus proteins may drive condensate formation {\em in vivo} in {\em E.\ coli}, due to their association with RNAPs which are in turn bound to {\em rrn operons}. 



In this work, we propose a generic physical model where we show that the formation of Nus condensates in {\em E.\ coli} leads to the subsequent colocalization of {\em rrn operons} due to the coarsening of the condensates. In our framework, weak interactions among the Nus proteins drive condensation through a PAC-like mechanism.  { Our model establishes  the mechanism by which the three-state dynamical behavior of Nus proteins reported in \cite{rnap_llps}, arises {\em in vivo}. }
Finally, we discuss the apparent discrepancy between the colocalization of {\em rrn operons} observed in fluorescence microscopy experiments and the corresponding Hi-C contact maps.





    



In the subsequent section of the manuscript, we first
describe our simulation model. There we outline all the interactions between
the chromosome-polymer and the proteins. Then we describe results from our
simulations in the ‘Results’ section. We finally conclude with a `Discussions' of
obtained results. 

\begin{figure}[!]
\hskip2.3cm
\includegraphics[width=0.45\columnwidth,angle=0]{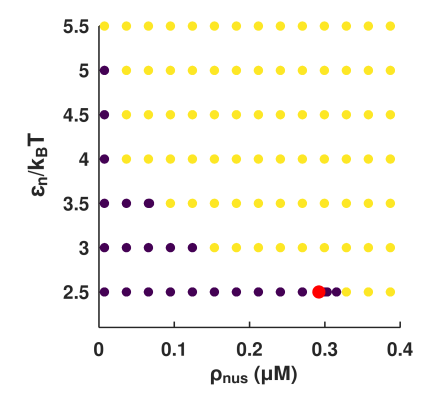}
\caption{\label{figbulk}
{\bf Bulk phase diagram of the Nus:} We display the bulk phase diagram of free Nus in the absence of the chromosome-polymer and bound Nus. 
The yellow region denotes the parameter regime where phase separation occurs, and the violet region corresponds to the homogeneous  phase. The red circle ($\epsilon_{n}/\mathrm{k_B T}=2.5$ and $\rho_{\rm nus}=0.29\, \mathrm{\mu M}$)  corresponds to the bulk state which we select for our simulations.}
\end{figure}

\section{ Simulation Model}

\noindent Our model comprises of four components: the chromosome polymer, the RNAP beads, the bound Nus beads and free Nus beads. In this section, we detail the various interactions among these components in our model. In our simulations we apply a Monte Carlo algorithm which is described further below\\

{\bf The chromosome:} We use a bead spring model of a flexible ring polymer with $460$ monomers representing the {\em E.\ coli} chromosome inside a cylindrical cell. 
Two neighboring monomers along the chain contour interact {\em via} the harmonic spring potential with energy  $V_H = \kappa (r - a)^2$, where $r$ is the distance between the monomers. The unit of length in our study is $a=1$, being the equilibrium length of the
harmonic-springs between two neighbouring monomers. The 
 spring constant $\kappa$ is $100\, \mathrm{k_B T/a^2}$. The interactions between  monomers are modeled 
by a Lennard-Jones (LJ) potential of the form
\begin{equation}
V_{LJ}(\epsilon,\sigma,r_c)=4\epsilon[(\sigma/r)^{12} - (\sigma/r)^{6}-(\sigma/r_c)^{12}+(\sigma/r_c)^{6}] 
\label{eq:LJ}
\end{equation}
 We consider the chromosome in good solvent and therefore implementing only repulsive, excluded volume (EV) interactions: $\epsilon=2.5\,\mathrm{k_B T}$, $\sigma=0.8a$ and  $r_c=2^{1/6}\sigma$.

The {\em E.\ coli} chromosome has $4.6\,\text{Mbp}$ of DNA, thus each monomer corresponds to $10$ $\text{kbp}$ of DNA.  We choose this degree of coarse-graining which corresponds to topologically distinct domains due to the presence of supercoiled loops \citep{supercoiled_loop}. Furthermore, in this study we are specifically interested in the spatial colocalization of {\em rrn operons} which are genetic sites spanning $\sim 5.5\,\text{kbp}$ of DNA which coincides with the degree of coarse-graining. Note that the Kuhn length of bare ds-DNA is  known to be comprised of  $300$ base pairs (bp) \citep{Phillips2012}. Thus, each monomer in our simulations corresponds to a length scale much greater than the Kuhn length and the polymer can be considered as fully flexible.\\

{\bf Sites of {\em rrn operons}:} Seven monomers along the polymer chain correspond to the seven sites of {\em rrn operons} in \textit{E. coli}. If the coarse-grained monomer having index $1$ is considered to be the {\em ori} (origin of replication), then the sites of {\em rrn operons} correspond to the monomers: $1, 3, 12, 25, 29, 121$ and $371$. The monomer indices of the {\em rrn operons} can be calculated based on the `genomic addresses' of the sites of {\em rrn operons} along the contour \citep{Gaal}.\\

{\bf Cylindrical confinement:}
Each coarse-grained monomer (of diameter $0.8a$) in our simulations comprises of $10\,\text{kbp}$ of DNA which represent topological domains formed by supercoiled loops which are further compacted by proteins. 
These domains  may roughly have a size of  $\approx 120\,\mathrm{nm}$ \citep{Ha2015,Pelletier2012,Jeon2017}. Thus, the unit of length in our simulations $a$, is given by $a\,\approx150\,\mathrm{nm}$. Since, the {\em E.\ coli} cell has a diameter of $1\,\mathrm{\mu m}$, the diameter of the cell in our simulation is $7a$. 
 The length of the cylinder is chosen to be $17.5a$. The aspect ratio of $1:2.5$ of the cylinder corresponds to the one observed {\em in-vivo} for an 
{\em E.\ coli} cell in slow growth conditions with a single mother chromosome \citep{Trueba869}. 
To simulate impenetrable cell walls
we reject any Monte Carlo trial move in which the center of the monomer or a protein attempts to occupy a position located outside the
cylinder.\\

{\bf RNAP proteins:} Proteins are considered as single beads on our level of coarse-graining. It is known that approximately $2000$ RNAPs are present in the {\em E.\ coli} cell (corresponding to slow growth conditions) {\em in vivo} and $50 \%$ of these RNAPs ($\approx 1000$) are actively involved in transcription \citep{rnap_number_proof}. Out of these actively transcribing RNAPs, $30\%$ of molecules
bind to sites of {\em rrn operons} \citep{Bremer2008}. Therefore, we consider $294$ RNAPs such that each {\em rrn operon} site is bound to  $42$ RNAPs. These RNAPs are bound to the sites of {\em rrn operons} by a harmonic interaction potential  $V_{rnap} = \kappa_r (r - a_r)^2$, where $r$ is the distance between a RNAP and a particular {\em rrn operon} site. The equilibrium length of the harmonic-spring between the RNAP and the {\em rrn operon} monomer is given by $a_r=\sigma_{r}/2 + \sigma/2$. The spring constant $\kappa_r$ is $100\, \mathrm{k_BT /a^2}$. The quantity $\sigma_{r}$ denotes the size of a RNAP and is given by $\sigma_{r}=0.1a$. This corresponds to a size of $15 \mathrm{nm}$ which roughly corresponds to the characteristic size of protein molecules \citep{Phillips2012}.  We do not consider the excluded volume of the RNAPs as in reality RNAPs are fixed and distributed along the {\em rrn operon} sequence. Therefore, the RNAP proteins are unlikely to interact with each other. In our model the RNAP play the role of linker elements between the coarse-grained operon and the bound Nus proteins, see Fig. \ref{fig1}. \\

{\bf Nus proteins:} Every RNAP binds one Nus, thus have $N_{bound}=294$  `bound' Nus in the cell. Each Nus bead is bound to a RNAP by a harmonic interaction potential  $V_{nus} = \kappa_n (r - a_n)^2$, where $r$ is the distance between a Nus bead and a particular RNAP. The equilibrium length of the harmonic-spring between the RNAP and Nus is given by $a_n=\sigma_{r}/2 + \sigma_{n}/2$. The spring constant $\kappa_n$ is $100\, \mathrm{k_B T/a^2}$. The quantity $\sigma_{n}$ denotes the size of Nus and is given by $\sigma_{n}=0.1a$. The bound Nus particles form a corona around the coarse-grain operon monomer, as sketched in Fig.\ref{fig1}.
In addition, we consider  a number ($N_{f}$) of `free' Nus which are not bound to RNAPs.\\

\begin{figure}[!]
\hskip0.9cm
\includegraphics[width=0.9\columnwidth,angle=0]{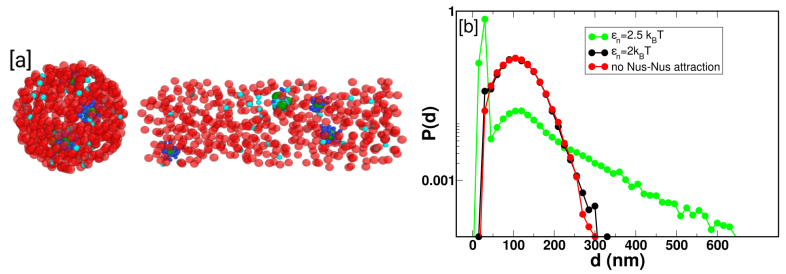}
\caption{\label{fig2}
{\bf Simulation snapshots and probability distribution of distance between neighbouring Nus:} (a) Snapshots (radial and the longitudinal view) from our simulations are displayed corresponding to $\epsilon_n=2.5\, \mathrm{k_B T}$ and $N_{f}=400$. The chromosome monomers are shown in red and monomers corresponding to the {\em rrn operons} are shown in green (enlarged). Bound Nus beads are shown in blue and free Nus shown in cyan (enlarged).  In (b) we display the probability distribution of the distance between between a free Nus bead and its nearest free Nus neighbour. We display the distributions for $\epsilon_n=2.5\, \mathrm{k_B T}$, $\epsilon_n=2.0\, \mathrm{k_B T}$ and when there are purely repulsive interactions between the Nus. The distribution has been computed with data collected after every $10,000$ MCS, from $1\times10^6$ MCS to $7.5\times 10^6$ MCS.  
}
\end{figure}

{\bf Interaction model of Nus:} 
The free Nus beads interact with each other {\em via} attractive Lennard Jones (LJ) interactions given by $V_{LJ}(\epsilon,\sigma,r_c)$ where $\epsilon=\epsilon_f$, $\sigma=\sigma_n$ and $r_c=2.5\sigma_n$, see Eq.\ref{eq:LJ}. The free Nus beads also interact with the bound Nus beads {\em via} attractive LJ interactions given by $V_{LJ}(\epsilon,\sigma,r_c)$ where $\epsilon=\epsilon_{bf}$, $\sigma=\sigma_n$ and $r_c=2.5\sigma_n$.

We choose equal interaction strengths between all Nus proteins, $\epsilon_{bf}=\epsilon_{f}=\epsilon_n$.  We deliberately choose $N_{f}=400$  (corresponding to $\rho_{\rm nus}=0.29\, \mathrm{\mu M}$) and $\epsilon_{n}=2.5\,\mathrm{k_B T}$, to ensure that the system is  outside the miscibility gap, see Fig.\ref{figbulk}.


 The `bound' Nus beads interact with each other only {\em via} the repulsive interaction given by $V_{LJ}(\epsilon,\sigma,r_c)$ where $\epsilon=\epsilon_{n}$, $\sigma=\sigma_n$ and $r_c=2^{1/6}\sigma_n$. This is because in our coarse-grained model there are $42$ `bound' Nus beads in close proximity to a single coarse-grained {\em rrn operon} monomer. In reality, they are bound to different parts  along the contour of the {\em rrn operon} sequence in the same way as RNAP and are thus unlikely to interact directly with each other. This is a consequence of our coarse-grained approach: on the one hand, each operon is represented as a single bead; on the other hand, we aim to capture the high effective functionality of operon regions arising from the large number of bound Nus proteins.\\

{\bf Monte Carlo Algorithm:} We apply a Monte Carlo dynamics in continuous space \citep{Binder_review}. In total there are $N$ beads in the simulation box (monomers, RNAPs and Nus). A Monte Carlo step (MCS) involves $N$ attempts to 
displace a bead in a random direction, with the maximum magnitude of the displacement in each direction being $0.1a$.  When attempting a displacement, the bead is chosen at random from  $N$ beads. The attempt is accepted if the Metropolis 
criterion is fulfilled thus ensuring convergence to the canonical distribution and detailed balance.

We first initialize the positions of the polymer monomers, the Nus beads, and the RNAP beads inside the cylindrical confinement. The system is then evolved using Monte Carlo simulations. To eliminate any initial spatial overlaps, we first simulate the system using a soft repulsive potential. The strength of the repulsive interaction is gradually increased from zero in discrete steps until it reaches its final value. In the simulations this is achieved by gradually increasing the value of $\sigma$ and $\sigma_n$.

We then estimate the time required for conformational relaxation of the polymer. To this end, we define a vector connecting two monomers separated by half the contour length of the ring, namely monomers $1$ and $230$.  We compute the autocorrelation of the corresponding unit vector and observe that the correlation decays to zero in approximately $1\times10^6$ Monte Carlo steps (MCS). This indicates that the polymer has equilibrated by $1\times10^6$ MCS. Thereafter, we implement the binding of RNAP molecules to the {\em rrn}-operon monomers, ensuring that each operon is bound to $42$ RNAPs. We then implement the binding of Nus beads to the RNAP molecules and ensure that each RNAP is bound to one Nus bead. After the binding of RNAPs and Nus, we switch on the attractive interactions between the Nus. We collect data only after the attractive interactions have been switched on.

{ In this work we study the dynamics of Nus proteins and the chromosome using Monte Carlo simulations using only local moves. It has been established
previously \citep{kikuchi1991,Romano2011} that Monte Carlo simulations with local moves, effectively reproduce diffusive dynamics 
of particles, at a coarse grained level. Thus, Monte Carlo simulations have been used widely on subjects ranging from phase
separation kinetics \citep{Singh2017} to polymer dynamics \citep{Binder_review,sharon}.}

\section{Results}
 {\bf Miscibility gap of Nus:} As mentioned in the introduction, Nus has been shown to phase separate {\em in vitro} at higher concentrations. Therefore, we first conduct simulations where there is only free Nus inside the cylinder in the absence of the chromosome polymer and RNAPs. We do this to identify the parameter regime in which Nus display a miscibility gap in our simulations. The results are shown in Fig.\ref{figbulk}. 

We vary both the concentration of Nus, $\rho_{\rm nus}$, as well as $\epsilon_{n}$ to construct the bulk phase diagram. For each point in the phase diagram we have computed the probability distribution of the distance between a Nus and its nearest neighbour. The bimodality of the probability distribution indicates the existence of two phases.  The yellow region  in Fig.\ref{figbulk} corresponds to the parameter regime where the system exhibits phase separation. In the violet region of the phase diagram, the system is in a homogeneous state.  In Fig.\ref{figbulk} we highlight a point in red which corresponds to the value of $\rho_{\rm nus}$ and $\epsilon_n$ that we use in our subsequent simulations. 
We have additionally checked that the addition of the chromosome polymer without attractive sites (bound Nus) does not change the phase behavior for the selected state.  \\

{\bf Formation of condensates by PAC-like mechanism:} We now establish that even outside the miscibility gap of free Nus, the system exhibits phase separation in the presence of the chromosome-polymer with bound Nus  for $\epsilon_n=2.5\, \mathrm{k_B T}$ and $N_f=400$ (corresponding to  $\rho_{\rm nus}=0.29\,\mathrm{\mu M}$, see red circle  in Fig.\ref{figbulk} ). Nus condensates are formed in close proximity to the sites of {\em rrn operons}. The high number of interacting sites (bound Nus) form a corona around the operon which acts as a (chemical) potential trap for the free Nus, see Fig.\ref{fig1} for illustration. As a result the concentration of the free Nus proteins is increased. When the local concentration exceeds the threshold for phase separation, a condensate nucleates in proximity to the sites of {\em rrn operons}. This mechanism is similar to the PAC mechanism proposed in \cite{Sommer_PAC}. Thus, the binding of Nus to RNAPs which are in turn bound to  {\em rrn operons} is crucial to induce phase separation. As these condensates coarsen with time, distant {\em rrn operon} sites are brought into proximity and become a part of the condensate.  In Fig.\ref{fig2}a we display a simulation snapshot of a case where condensate formation has occurred. The monomers corresponding to the {\em rrn operons} are shown in green. These monomers have been enlarged for aid of visualization. The bound Nus are shown in blue and have been enlarged for aid of visualization, RNAPs are not shown. In Fig.\ref{fig2}b we display the probability distribution of the distance between a free Nus and its nearest neighbour. When $\epsilon_n=2.5\, \mathrm{k_B T}$, a bimodal distribution with two separated peaks is seen, which indicates that two phases have been formed.

\begin{figure}[!]
\hskip0.8cm
\includegraphics[width=0.9\columnwidth]{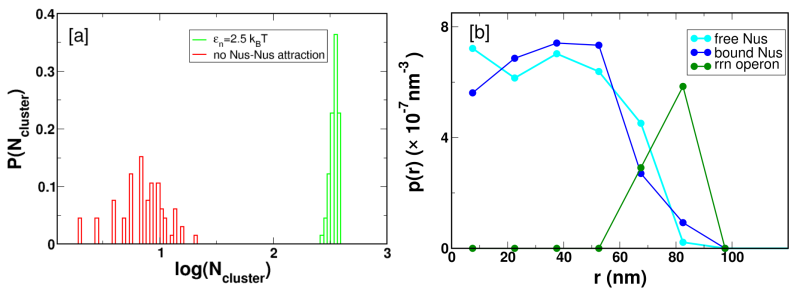}
\caption{\label{fig3}
{\bf Number of Nus particles in cluster and spatial distribution of components:}
In (a)  we display the probability distribution of the number of free Nus particles in the largest cluster ($N_{cluster}$), formed during a simulation run. We display data for $\epsilon_n=2.5\, \mathrm{k_B T}$ and for the case when there are only repulsive interactions among Nus. In (b) we display the radial probability densities $p(r)$ of the various components as a function of the distance $r$ from the center of mass (COM) of the cluster. The quantity $p(r)\cdot4\pi r^{2}\,dr$ gives the probability of finding a particle in a spherical shell between $r$ and $r+dr$. The radial bin size used in the calculation is $\delta r=15\,\mathrm{nm}$. The distributions shown in (a) and (b) have been computed with data collected after every $10,000$ MCS from $1\times10^6$ MCS to $7.5\times 10^6$ MCS. 
 }
\end{figure}

When the strength of the attractive interactions ($\epsilon_n$) is reduced significantly below  $2.5\, \mathrm{k_B T}$ then phase separation no longer occurs and the distribution is single peaked such as shown for for $\epsilon_n=2\, \mathrm{k_B T}$.  In Fig.\ref{fig2}b, we also display the case where there are only repulsive interactions between the Nus beads for comparison. From these results we can conclude that the mean distance between non-condensed free Nus is about $110\,\mathrm{nm}$.\\


{\bf Spatial distribution of components inside the condensate:} In Fig.\ref{fig3}a we display the probability distribution of the number of free Nus particles in the largest cluster ($N_{cluster}$), formed during a simulation run. We display data for $\epsilon_n=2.5\, \mathrm{k_B T}$ and for the case when there are only repulsive interactions among Nus. To compute the number of free Nus particles in the largest cluster, we first used a distance-based connectivity analysis to identify all the members of the largest cluster. We aim to identify all the free and bound Nus particles as well as the {\em rrn operon} monomers which may be a part of the cluster. For each simulation frame (collected after every $10^4$ MCS), pairwise distances between the particles  were computed, and two particles were considered connected if their separation was smaller than a prescribed cutoff distance of $75\,\mathrm{nm}$. We use a cutoff distance of size of the operons (having a radius of $60\,\mathrm{nm}$) which are a part of the cluster. Also, the chosen cutoff distance is still considerably lower than the distance between neighbouring Nus particles in the dilute phase which is approximately $110\,\mathrm{nm}$, as displayed in Fig.\ref{fig2}b. Clusters were then identified  using an iterative procedure: starting from an unassigned particle, all neighboring particles within the cutoff were recursively added, followed by neighbors of those neighbors, until no further particles could be included. After all members of the largest cluster were identified, the number of free Nus particles in the cluster,$N_{cluster}$,  was calculated. 

In Fig.\ref{fig3}a we display the distribution of $N_{cluster}$ for two values of the interaction parameters between Nus. Without attractive interactions only small clusters are statistically formed which corresponds to the non-condensed phase. For the case of attractive interactions the largest cluster corresponds to a condensed state which incorporates several hundreds of Nus-particles. 

In Fig.\ref{fig3}b we display the radial probability densities $p(r)$ of the various components as a function of the distance $r$ from the center of mass (COM) of the largest cluster. It can be inferred from Fig.\ref{fig3}b that the {\em rrn operons} occupy the periphery of the cluster with higher likelihood, while the free and bound Nus are preferentially found closer to the center of the cluster, but have broad distributions. 
\\

\begin{figure}[!]

\hskip0.6cm
 \includegraphics[width=0.9\columnwidth,angle=0]{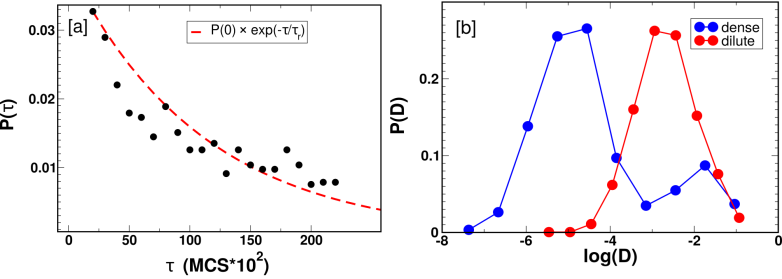}
\caption{\label{fig4}
{\bf Dynamics of Nus in the two phases:} In (a) we display the probability distribution of free Nus to be inside the condensate for time $\tau$ (MCS). The mean residence time $\tau_r$, is extracted by fitting an exponential function: $\mathrm{P(0)}\cdot \exp(-\tau/\tau_r)$  to the curve. The mean residence time is $\tau_r \simeq  11,000$ MCS. 
In  (b) we display the distributions of diffusion constants of free Nus in the dilute and the dense phase respectively. The most probable diffusion constant value in the two phases differ by approximately two orders of magnitude.}
\end{figure}

{\bf Dynamics of Nus in the two phases:} We now investigate the dynamics of the free Nus inside the condensate. We emphasize that the condensates are dynamic in the sense that the dilute and the condensed phases exchange components (free Nus beads) frequently. A relevant question to ask is: what is the mean residence time of the free Nus bead inside the cluster? To that end, we calculate the probability distribution of a free Nus bead to be in the cluster for time $\tau$ measured in MCS, refer Fig.\ref{fig4}a. To calculate the residence time we calculate how how long a free Nus bead remains continuously within the largest cluster over successive simulation frames.  Therefore, in Fig.\ref{fig4}a we display the probability distribution of  a free Nus bead to be a part of the cluster for time $\tau$. We further fit an exponential function given by $\mathrm{P(0)}\cdot \exp(-\tau/\tau_r)$ to the curve. The quantity $\tau_r$ is the mean residence time of a free Nus bead to be inside the cluster. We find that $\tau_r\simeq 11,000$ MCS which is much shorter than the length of a typical simulation run ($\approx 1\times10^7$ MCS). Thus, free Nus dynamically diffuse in and out of the condensate during the simulation run.  

To further establish a correspondence with experiments in \cite{rnap_llps}, we calculate the mean square displacement (MSD) of free Nus as a function of time (MCS).  The diffusion constant for free Nus in the dense state, was estimated during the time interval in which a free Nus bead resides within the largest cluster. We record the positions of free Nus particles, every $100$ MCS. Continuous residence durations were then determined by identifying sequences of consecutive frames in which the bead remained in the largest cluster.  For each residence episode longer than one frame,  we first identify the frame where the particle is part of the cluster for the first time. This corresponds to time $t=t_{1}$. We then identify the last frame where the particle is part of the cluster. This corresponds to time $t=t_{2}$. The squared displacement over this interval given by $\Delta r^2=  (r(t_{2}) -r(t_{1}))^2$ was computed. The diffusion constant was then obtained as $D=\Delta r^2/6\Delta t$ where $\Delta t=t_{2}-t_{1}$. This analysis was then repeated for different residence episodes for the same Nus bead. We do this for all the Nus particles that are ever part of the largest cluster during the entire simulation run. Thus, the diffusion constants obtained can then be used to calculate the probability distribution displayed in Fig.\ref{fig4}b. Similarly the diffusion constant of the free Nus in the dilute phase, can be estimated during the duration in which a free Nus bead resides outside the cluster. Thus, the probability distribution of diffusion constants of Nus in the dilute phase as shown in Fig.\ref{fig4}b was obtained.

We note that the most probable diffusion constant value in the two phases differ by approximately two orders of magnitude. This is similar to what was reported in \cite{rnap_llps}. Note that the distribution of the dense phase in Fig.\ref{fig4}b is  broader than the distribution of the dilute phase. This is because in the dense phase, there are some free Nus beads which are found close to the surface of the cluster. These beads are then part of the cluster for a short times before they diffuse into the dilute phase. The dynamics of these beads in those short durations are then relatively unaffected by the dense environment in the interior of the condensate. Therefore, these beads have a diffusion coefficient similar to what is seen in the bulk phase.   \\  

{\bf Colocalization of {\em rrn operons} and contact maps:} In  the work of \cite{Gaal} the authors showed that any pair of {\em rrn operons} are statistically closer to each other, than any other two genomic segments separated by equivalent genomic distance, i.e. {\em rrn operons} are colocalised. To reconcile the experimental observations using our model, we display the probability distribution of the pairwise distance between {\em rrn operons} in Fig.\ref{fig5}a. It is seen that when the system shows phase separation and condensates are formed for $\epsilon_n=2.5\, \mathrm{k_B T}$, the probability distribution of the pairwise distances between {\em rrn operons} shows a pronounced peak at small values corresponding to the distance between those {\em rrn operon} pairs that get spatially colocalized. The second maximum coincides approximately with the peak of the distribution of non-condensed {\em rrn operons}. Not all operons get spatially colocalized in a single cluster during a simulation run. This causes a multi-modal distribution where the additional maxima indicate the spatial separation of the different clusters.

In Fig.\ref{fig5}b we display the contact map of the chromosome-polymer. If the distance between any two monomers is less than $210\, \mathrm{nm}$  then the monomers are deemed to be in contact. For any two monomers having indices $i$ and $j$, the color of the contact map at the coordinate (i,j) indicates the probability of contact of monomers $i$ and $j$. The bright diagonal that can be seen in the contact map arises from contacts of monomers with their adjacent neighbours along the contour. Bright spots away from the diagonal correspond to the sites of {\em rrn operons} which come into contact as the condensate gets formed. In Fig.\ref{fig5}b, it is seen that the {\em rrn operon} monomers having indices $3$ and $12$ are in contact. The contact map shown in Fig.\ref{fig5}b has been computed by averaging from  $1\times10^6$ MCS to $4\times10^6$ MCS. At later times, as the clusters coarsen, other {\em rrn operon} monomers come into contact. In Fig.\ref{fig5}c we display the contact map corresponding to later times, which has been computed by averaging from  $5\times10^6$ MCS to $7.5\times10^6$ MCS. It is seen that the monomers $3$ and $12$ have continued to be  in contact. In addition, the {\em rrn operon} monomer having index $121$ forms contacts with monomers $3$ and $12$. It may be  expected that at long times all the sites of {\em rrn operons} would be spatially colocalized since the entropy effort to form additional loops scales with the logarithm of the loop length and is thus of order $k_\mathrm{B} T$ and coalescence of the individual clusters is a very slow process. {It is worth noting that a theoretical estimate of the Rouse relaxation time for the {\em E. coli} chromosome is of the order of a few hours. Recent experimental evidence has shown that {\em E. coli} loci exhibit subdiffusive motion with an exponent of $0.4$ \citep{Weber2010}, rather than the value of $0.5$ expected for a Rouse chain, at timescales shorter than the Rouse time. This suggests that the chromosome may relax even more slowly than the estimated Rouse time of a few hours. In contrast, the typical cell cycle of {\em E. coli} lasts approximately $20-50$ minutes \citep{Phillips2012}. Thus, the chromosome relaxation time is substantially longer than the cell-cycle duration. It is therefore plausible that not all operons have sufficient time to spatially colocalize {\em in vivo}. This may lead to a significant reduction in the ensemble-averaged contact probability between operon pairs and may provide an additional explanation for the low contact probabilities of operon pairs observed in experimentally obtained Hi-C maps.
}

{ While our model provides a rational explanation for the formation of Nus condensates induced by Nus proteins bound to the {\em rrn operons} and successfully explains the experimental observations reported in the work of ~\cite{Gaal}, the situation in living cells is more dynamic: neither Nus proteins nor RNAP are permanently bound to the operon.
In reality RNAPs frequently disassociate from chromosomal segments even as transcription occurs \citep{rnap_residence,Phillips2012} at the timescale of $1-100$ s. In our coarse-grained model this means that bound Nus particles leave the corona, see Fig.\ref{fig1}, and others can be attached to newly bound RNAPs. We have therefore extended the simulation model by implementing stochastic unbinding-rebinding of RNAPs and Nus. For details see section-I and II of the Supplementary Information (SI). We find that the incorporation of stochastic unbinding-binding kinetics do not lead to a qualitative change of the result reported for the permanently bound components. We observe either formation of stable condensates or there is no formation of condensates at all, depending on the mean lifetime of a RNAP-operon bond. Obviously, if binding and unbinding events occur too rapidly, the ability of the bound Nus proteins to mark the {\em rrn operons} and thereby nucleate condensates of free Nus proteins is lost. Indeed, our simulations reveal a sharp transition in condensate formation as a function of the mean lifetime of the RNAP-operon bonds, denoted by $\tau_{RO}$. In particular our model predicts that if $\tau_{RO} \gtrsim 0.06s$  then condensate formation is likely to occur. It is known that the unbinding of RNAPs occur {\em in vivo} at a timescale of $1-100\ s$ \citep{Phillips2012,rnap_residence}. Thus, it is plausible that condensate formation does indeed occur {\em in vivo}.

\begin{figure*}[!]
\includegraphics[width=\columnwidth,angle=0]{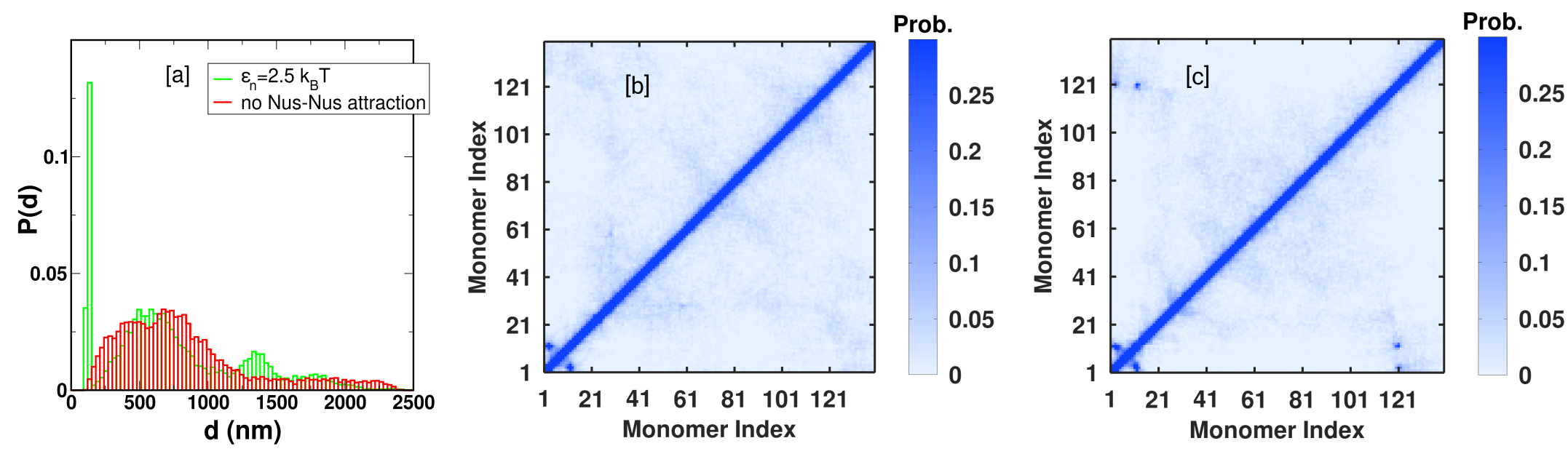}
\caption{\label{fig5}
{\bf Colocalization of {\em rrn operons}:} In (a) we display the probability distribution of the pairwise distance between {\em rrn operons} for  $\epsilon_n=2.5\, \mathrm{k_B T}$ and for only repulsive interactions between Nus, computed over $1\times10^5$ to  $7.5\times10^6$ MCS.  In (b) we display the contact map of the chromosome-polymer, which has been computed by averaging from  $1\times10^6$ MCS to $4\times10^6$ MCS with data collected after every $10,000$ MCS.  The monomers having indices $3$ and $12$ show high contact probabilities. 
In (c) we display the contact map at later times, averaging from  $5\times10^6$ MCS to $7.5\times10^6$ MCS.  The monomers having indices $3$, $12$ and $121$ show high contact probabilities.
}
\end{figure*}

In Fig.1 of SI, we also present results where binding-unbinding kinetics is not incorporated, but the number of RNAPs (and Nus) bound to the operons is systematically varied. The correponding discussion is presented in Section-II of SI. We find that a minimum of $26$ Nus proteins associated to each operon site is required for condensate formation. The number of RNAPs associated with an {\em rrn operon} is estimated to vary between $\approx20-72$ {\em in vivo}, depending on the growth condition \citep{achilles}. During fast growth, a relatively larger number of RNAPs are engaged in transcription at the {\em rrn operons}, and RNAP clustering has been observed. Our results therefore suggest that the increased RNAP occupancy at the sites of {\em rrn operons} under fast-growth can lead to the formation of condensates, due to a PAC-like mechanism as outlined here.\\

It is expected that Nus proteins interact with the nascent RNA as well as other bio-molecules involved in transcription. Such interactions are likely to modulate the effective interactions between Nus proteins over time. To represent these effects at a phenomenological coarse-grained level, we introduce temporal fluctuations in the strength of the Lennard-Jones interactions ($\epsilon_n$) among the Nus beads, without incorporating the binding-unbinding kinetics of the proteins. The details of the implementation as well as the results have been presented in Section-III of SI. Such active (non-thermal) temporal fluctuations indeed show that the condensates form and dissolve transiently, leading to colocalization of {\em rrn operons} over short times ($\approx 1s$). However, the probability distribution of the pairwise distance between the {\em rrn operons} still shows a peak at small distances. Thus, our model with the additional incorporation of `active noise' in the interactions among the Nus can suggest a potential reconciliation of results obtained from imaging experiments \citep{Gaal} as well as Hi-C experiments \citep{Lioy2018}, where co-localization of rrn operons have not been detected yet. As stated before, a crucial step in Hi-C experiments is the formaldehyde-mediated cross-linking of spatially proximal DNA segments. Since this cross-linking process occurs on a timescale of the order of 10 minutes \citep{hic_original,Polovnikov2026}, transient colocalization of the {\em rrn operons} is unlikely to be efficiently captured. If the {\em rrn operons} remain colocalized only for much shorter periods, cross-links between them will rarely form during sample preparation, providing a plausible explanation for the absence of an {\em rrn operon} colocalization signature in Hi-C experiments. { The absence of the signature of {\em rrn operon} colocalization in Hi-C experiments may also be due to various other experimental factors, as already mentioned previously.}}

\section{Discussion}

 In this work, we establish a model by which Nus-proteins form clusters even outside the bulk miscibility gap via a PAC-like mechanism, which can explain the colocalization of {\em rrn operons} observed in recent experiments. We propose a coarse-grained model in which topologically distinct domains of the bacterial DNA on the order of 10 kbp are mapped onto effective beads, resulting in a cyclic flexible polymer of 460 repeat units confined within a cylinder representing the {\em E.\ coli} cell. The {\em rrn operons} are represented by special beads to which Nus proteins are bound via RNAP proteins. In essence, the operon units display a corona of Nus proteins that act as highly functional interaction sites with respect to freely diffusing Nus proteins in the cytoplasm. We show that this scenario leads to phase separation of free Nus together with the Nus-decorated {\em rrn operons}, thus explaining the experimentally observed colocalization of {\em rrn operons}. 

Further evidence for the formation of a dense Nus-phase comes from experiments measuring the diffusion dynamics of Nus. In addition to bound Nus, two diffusive populations have been identified, corresponding to slow diffusion (in an apparently denser environment) and normal diffusion in the cytoplasm. Our simple polymer-physics-based model reproduces this behavior as a consequence of the partitioning of Nus into a condensed phase and a dilute phase. 

{ A puzzling aspect is that colocalization of the {\em rrn operons} has not yet been observed in Hi-C maps. {One of the} possible explanations could be that operon clusters reorganize on time scales much shorter than those required for the cross-linking of spatially proximal DNA sequences during Hi-C sample preparation. Although our coarse-grained model cannot make quantitative predictions for real time scales, the operon clusters remain rather stable within the parameter regime considered here.

To suggest one possible reconciliation of the apparent discrepancy between the results obtained from live-cell single-molecule tracking and Hi-C, we incorporated active (non-thermal) fluctuations into our model by allowing the interaction strength between Nus proteins to fluctuate on a time scale of approximately 1s, comparable to the typical attachment and detachment time of RNAP. These actively driven fluctuations enhance cluster dynamics and suppress further coarsening at long times. As a consequence, averaging the contact map over extended time periods corresponding to sampling a large population of cells—progressively diminishes cluster-specific contact signatures, whereas spatial correlation functions continue to reveal the colocalization of the {\em rrn operons}.}

In conclusion, we show that phase separation of Nus proteins assisted by Nus-decorated {\em rrn operons} can rationalize various, partly seemingly contradictory, experimental observations in a semi-quantitative manner. An interesting question is to what extent the colocalization of {\em rrn operons} supports efficient ribosome production in bacterial cells. Drawing an analogy to eukaryotic cells, where the ribosome “factory” is the complex multiphase nucleolar condensate, one may speculate that Nus clusters play a similar role or at least form part of an analogous scaffold in bacteria. In this context, our model still lacks the aspect of rRNA extrusion and subsequent biochemical processing of ribosomal components. In particular, the constant flux of nascent rRNA from operon sites may influence condensate properties and, in particular, limit coarsening, similar to our ad hoc approach of fluctuating interactions among Nus proteins. Such effects could be incorporated into the Monte Carlo algorithm and will be considered in future work.

 \section{Acknowledgments}
 \noindent This work was supported by the Deutsche Forschungsgemeinschaft (DFG, German Research Foundation) through the Research Training Group GRK 3120. J.U.S acknowledges support by the DFG under the grant SO 277/25 and by Germany’s Excellence Strategy-EXC2068-390729961-Cluster of Excellence Physics of Life.\\

\section{Author Contributions}
\noindent J.U.S.: Conceptualization, model development, writing
D.M.: Model development, simulations, data analysis, visualization, validation, writing. All authors contributed to discussion and approved the final manuscript.\\



\begin{thebibliography}{45}
\providecommand{\natexlab}[1]{#1}
\providecommand{\urlprefix}{}
\providecommand{\doiprefix}{doi: }

\bibitem[{Aierken et~al.(2026)Dilimulati Aierken and Sebastian Aland and
  Stefano Bo and Steven Boeynaems and Danfeng Cai and Serena Carra and Lindsay
  B. Case and Hue Sun Chan and Jorge R. Espinosa and Trevor K. GrandPre and
  Alexander Y. Grosberg and Ivar S. Haugerud and William M. Jacobs and Jerelle
  A. Joseph and Frank Jülicher and Kurt Kremer and Guido Kusters and Liedewij
  Laan and Keren Lasker and Katrin S. Laxhuber and Hyun O. Lee and Kathy F. Liu
  and Dimple Notani and Yicheng Qiang and Paul Robustelli and Leonor Saiz and
  Omar A. Saleh and Helmut Schiessel and Jeremy Schmit and Meng Shen and
  Krishna Shrinivas and Antonia Statt and Andres R. Tejedor and Tatjana Trcek
  and Christoph A. Weber and Stephanie C. Weber and Ned S. Wingreen and
  Huaiying Zhang and Yaojun Zhang and Huan Xiang Zhou and David
  Zwicker}]{roadmap}
\textbf{\color{eLifeMediumGrey} Aierken D}, Aland S, Bo S, Boeynaems S, Cai D,
  Carra S, Case LB, Chan HS, Espinosa JR, GrandPre TK, Grosberg AY, Haugerud
  IS, Jacobs WM, Joseph JA, Jülicher F, Kremer K, Kusters G, Laan L, Lasker K,
  Laxhuber KS, et~al., Roadmap for Condensates in Cell Biology; 2026.
\newblock \urlprefix\url{https://arxiv.org/abs/2601.03677}.

\bibitem[{Alberti and Dormann(2019)Alberti, Simon and Dormann,
  Dorothee}]{Alberti2019}
\textbf{\color{eLifeMediumGrey} Alberti S}, Dormann D.
\newblock Liquid–Liquid Phase Separation in Disease.
\newblock Annual Review of Genetics.  2019 Dec; 53(1):171–194.
\newblock
  \urlprefix\url{http://dx.doi.org/10.1146/annurev-genet-112618-043527},
  \href{10.1146/annurev-genet-112618-043527}{\doiprefix
  \detokenize{10.1146/annurev-genet-112618-043527}}.

\bibitem[{Bakshi et~al.(2012)Bakshi, Somenath and Siryaporn, Albert and
  Goulian, Mark and Weisshaar, James C.}]{rnap_number_proof}
\textbf{\color{eLifeMediumGrey} Bakshi S}, Siryaporn A, Goulian M, Weisshaar
  JC.
\newblock Superresolution imaging of ribosomes and RNA polymerase in live
  Escherichia coli cells.
\newblock Molecular Microbiology.  2012 May; 85(1):21–38.
\newblock \urlprefix\url{http://dx.doi.org/10.1111/j.1365-2958.2012.08081.x},
  \href{10.1111/j.1365-2958.2012.08081.x}{\doiprefix
  \detokenize{10.1111/j.1365-2958.2012.08081.x}}.

\bibitem[{Binder and Paul(1997)Binder, K. and Paul, W.}]{Binder_review}
\textbf{\color{eLifeMediumGrey} Binder K}, Paul W.
\newblock Monte Carlo simulations of polymer dynamics: Recent advances.
\newblock Journal of Polymer Science Part B: Polymer Physics.  1997;
  35(1):1--31.
\newblock
  \urlprefix\url{https://onlinelibrary.wiley.com/doi/abs/10.1002/%28SICI%291099-0488%2819970115%2935%3A1%3C1%3A%3AAID-POLB1%3E3.0.CO%3B2-%23},
  \href{https://doi.org/10.1002/(SICI)1099-0488(19970115)35:1<1::AID-POLB1>3.0.CO;2-\#}{\doiprefix
  \detokenize{https://doi.org/10.1002/(SICI)1099-0488(19970115)35:1<1::AID-POLB1>3.0.CO;2-\#}}.

\bibitem[{Brangwynne et~al.(2009)Brangwynne, Clifford P. and Eckmann, Christian
  R. and Courson, David S. and Rybarska, Agata and Hoege, Carsten and
  Gharakhani, J\"{o}bin and J\"{u}licher, Frank and Hyman, Anthony
  A.}]{Brangwynne2009}
\textbf{\color{eLifeMediumGrey} Brangwynne CP}, Eckmann CR, Courson DS,
  Rybarska A, Hoege C, Gharakhani J, J\"{u}licher F, Hyman AA.
\newblock Germline {P Granules} Are Liquid Droplets That Localize by Controlled
  Dissolution/Condensation.
\newblock Science.  2009 Jun; 324(5935):1729–1732.
\newblock \urlprefix\url{http://dx.doi.org/10.1126/science.1172046},
  \href{10.1126/science.1172046}{\doiprefix
  \detokenize{10.1126/science.1172046}}.

\bibitem[{Bremer and Dennis(2008)Bremer, Hans and Dennis, Patrick
  P.}]{Bremer2008}
\textbf{\color{eLifeMediumGrey} Bremer H}, Dennis PP.
\newblock Modulation of Chemical Composition and Other Parameters of the Cell
  at Different Exponential Growth Rates.
\newblock EcoSal Plus.  2008 Jan; 3(1).
\newblock \urlprefix\url{http://dx.doi.org/10.1128/ecosal.5.2.3},
  \href{10.1128/ecosal.5.2.3}{\doiprefix \detokenize{10.1128/ecosal.5.2.3}}.

\bibitem[{Caragine et~al.(2019)Caragine, Christina M and Haley, Shannon C and
  Zidovska, Alexandra}]{nucleolus2}
\textbf{\color{eLifeMediumGrey} Caragine CM}, Haley SC, Zidovska A.
\newblock Nucleolar dynamics and interactions with nucleoplasm in living cells.
\newblock eLife.  2019 nov; 8:e47533.
\newblock \urlprefix\url{https://doi.org/10.7554/eLife.47533},
  \href{10.7554/eLife.47533}{\doiprefix \detokenize{10.7554/eLife.47533}}.

\bibitem[{Cass et~al.(2016)Julie A. Cass and Nathan J. Kuwada and Beth
  Traxler and Paul A. Wiggins}]{Cass2016}
\textbf{\color{eLifeMediumGrey} Cass J}, Kuwada N, Traxler B, Wiggins P.
\newblock Escherichia coli Chromosomal Loci Segregate from Midcell with
  Universal Dynamics.
\newblock Biophysical Journal.  2016; 110(12):2597--2609.
\newblock
  \urlprefix\url{https://www.sciencedirect.com/science/article/pii/S0006349516302727},
  \href{https://doi.org/10.1016/j.bpj.2016.04.046}{\doiprefix
  \detokenize{https://doi.org/10.1016/j.bpj.2016.04.046}}.

\bibitem[{Fan et~al.(2023)Fan, Jun and El Sayyed, Hafez and Pambos, Oliver J
  and Stracy, Mathew and Kyropoulos, Jingwen and Kapanidis, Achillefs
  N}]{achilles}
\textbf{\color{eLifeMediumGrey} Fan J}, El Sayyed H, Pambos OJ, Stracy M,
  Kyropoulos J, Kapanidis AN.
\newblock {RNA polymerase redistribution supports growth in E. coli strains
  with a minimal number of rRNA operons}.
\newblock Nucleic Acids Research.  2023 06; 51(15):8085--8101.
\newblock \urlprefix\url{https://doi.org/10.1093/nar/gkad511},
  \href{10.1093/nar/gkad511}{\doiprefix \detokenize{10.1093/nar/gkad511}}.

\bibitem[{Feric et~al.(2016)Feric, Marina and Vaidya, Nilesh and Harmon, Tyler
  S. and Mitrea, Diana M. and Zhu, Lian and Richardson, Tiffany M. and
  Kriwacki, Richard W. and Pappu, Rohit V. and Brangwynne, Clifford
  P.}]{Feric2016}
\textbf{\color{eLifeMediumGrey} Feric M}, Vaidya N, Harmon TS, Mitrea DM, Zhu
  L, Richardson TM, Kriwacki RW, Pappu RV, Brangwynne CP.
\newblock Coexisting Liquid Phases Underlie Nucleolar Subcompartments.
\newblock Cell.  2016 Jun; 165(7):1686–1697.
\newblock \urlprefix\url{http://dx.doi.org/10.1016/j.cell.2016.04.047},
  \href{10.1016/j.cell.2016.04.047}{\doiprefix
  \detokenize{10.1016/j.cell.2016.04.047}}.

\bibitem[{Gaal et~al.(2016)Tamas Gaal and Benjamin P. Bratton and Patricia
  Sanchez-Vazquez and Alexander Sliwicki and Kristine Sliwicki and Andrew Vegel
  and Rachel Pannu and Richard L. Gourse}]{Gaal}
\textbf{\color{eLifeMediumGrey} Gaal T}, Bratton BP, Sanchez-Vazquez P,
  Sliwicki A, Sliwicki K, Vegel A, Pannu R, Gourse RL.
\newblock Colocalization of distant chromosomal loci in space in {E. coli}: a
  bacterial nucleolus.
\newblock Genes and Development.  2016 Oct; 30(20):2272--2285.
\newblock \urlprefix\url{https://doi.org/10.1101/gad.290312.116},
  \href{10.1101/gad.290312.116}{\doiprefix
  \detokenize{10.1101/gad.290312.116}}.

\bibitem[{Gao et~al.(2023)Gao, Guoming and Sumrall, Emily S. and Pitchiaya,
  Sethuramasundaram and Bitzer, Markus and Alberti, Simon and Walter, Nils
  G.}]{Gao2023}
\textbf{\color{eLifeMediumGrey} Gao G}, Sumrall ES, Pitchiaya S, Bitzer M,
  Alberti S, Walter NG.
\newblock Biomolecular condensates in kidney physiology and disease.
\newblock Nature Reviews Nephrology.  2023 Sep; 19(12):756–770.
\newblock \urlprefix\url{http://dx.doi.org/10.1038/s41581-023-00767-0},
  \href{10.1038/s41581-023-00767-0}{\doiprefix
  \detokenize{10.1038/s41581-023-00767-0}}.

\bibitem[{Glotzer et~al.(1994)Glotzer, Sharon C. and Stauffer, Dietrich and
  Jan, Naeem}]{sharon}
\textbf{\color{eLifeMediumGrey} Glotzer SC}, Stauffer D, Jan N.
\newblock Monte Carlo simulations of phase separation in chemically reactive
  binary mixtures.
\newblock Physical Review Letters.  1994 Jun; 72:4109--4112.
\newblock \urlprefix\url{https://link.aps.org/doi/10.1103/PhysRevLett.72.4109},
  \href{10.1103/PhysRevLett.72.4109}{\doiprefix
  \detokenize{10.1103/PhysRevLett.72.4109}}.

\bibitem[{Ha and Jung(2015)Ha, Bae-Yeun and Jung, Youngkyun}]{Ha2015}
\textbf{\color{eLifeMediumGrey} Ha BY}, Jung Y.
\newblock Polymers under confinement: single polymers, how they interact, and
  as model chromosomes.
\newblock Soft Matter.  2015; 11(12):2333–2352.
\newblock \urlprefix\url{http://dx.doi.org/10.1039/c4sm02734e},
  \href{10.1039/c4sm02734e}{\doiprefix \detokenize{10.1039/c4sm02734e}}.

\bibitem[{Harmon et~al.(2020)Harmon, Tyler S. and Mitrea, Diana M. and
  Kriwacki, Richard and Julicher, Frank}]{Harmon2020}
\textbf{\color{eLifeMediumGrey} Harmon TS}, Mitrea DM, Kriwacki R, Julicher F.
\newblock Converting Stochastic Assembly into an Assembly Line: Non-Equilibrium
  Droplet Dynamics Assist Ribosome Formation.
\newblock Biophysical Journal.  2020 Feb; 118(3):370a.
\newblock \urlprefix\url{http://dx.doi.org/10.1016/j.bpj.2019.11.2121},
  \href{10.1016/j.bpj.2019.11.2121}{\doiprefix
  \detokenize{10.1016/j.bpj.2019.11.2121}}.

\bibitem[{Ho and Harshey(2025)Ho, Khang and Harshey, Rasika M.}]{Ho2025}
\textbf{\color{eLifeMediumGrey} Ho K}, Harshey RM.
\newblock Membrane-associated $\sigma$ factors disrupt rRNA operon clustering in
  Escherichia coli.
\newblock PLOS Biology.  2025 Apr; 23(4):e3003113.
\newblock \urlprefix\url{http://dx.doi.org/10.1371/journal.pbio.3003113},
  \href{10.1371/journal.pbio.3003113}{\doiprefix
  \detokenize{10.1371/journal.pbio.3003113}}.

\bibitem[{Ho et~al.(2025)Ho, Khang and Royzenblat, Sonya K. and Wilkins, Brady
  and Harshey, Rasika and Freddolino, Lydia}]{mu1}
\textbf{\color{eLifeMediumGrey} Ho K}, Royzenblat SK, Wilkins B, Harshey R,
  Freddolino L.
\newblock In: High-Throughput Mapping of Chromosomal Conformations in E. coli
  Under Physiological Conditions Using Massively Multiplexed Mu Transposition
  Springer US; 2025. p. 125–146.
\newblock \urlprefix\url{http://dx.doi.org/10.1007/978-1-0716-3930-6_30},
  \href{10.1007/978-1-0716-3930-6_30}{\doiprefix
  \detokenize{10.1007/978-1-0716-3930-6_30}}.

\bibitem[{Hodgins et~al.(2025)Hodgins, Lydia and Parmar, Baljyot Singh and
  Reyes-Lamothe, Rodrigo and Weber, Stephanie C.}]{Hodgins2025}
\textbf{\color{eLifeMediumGrey} Hodgins L}, Parmar BS, Reyes-Lamothe R, Weber
  SC.
\newblock Size Matters: A Biophysical Perspective on Biomolecular Condensates
  in Bacteria.
\newblock Annual Review of Biophysics.  2025 Dec;
  \urlprefix\url{http://dx.doi.org/10.1146/annurev-biophys-021424-010232},
  \href{10.1146/annurev-biophys-021424-010232}{\doiprefix
  \detokenize{10.1146/annurev-biophys-021424-010232}}.

\bibitem[{Hyman et~al.(2014)Hyman, Anthony A. and Weber, Christoph A. and
  J\"{u}licher, Frank}]{Hyman2014}
\textbf{\color{eLifeMediumGrey} Hyman AA}, Weber CA, J\"{u}licher F.
\newblock Liquid-Liquid Phase Separation in Biology.
\newblock Annual Review of Cell and Developmental Biology.  2014 Oct;
  30(1):39–58.
\newblock
  \urlprefix\url{http://dx.doi.org/10.1146/annurev-cellbio-100913-013325},
  \href{10.1146/annurev-cellbio-100913-013325}{\doiprefix
  \detokenize{10.1146/annurev-cellbio-100913-013325}}.

\bibitem[{Jeon et~al.(2017)Jeon, Chanil and Jung, Youngkyun and Ha,
  Bae-Yeun}]{Jeon2017}
\textbf{\color{eLifeMediumGrey} Jeon C}, Jung Y, Ha BY.
\newblock A ring-polymer model shows how macromolecular crowding controls
  chromosome-arm organization in Escherichia coli.
\newblock Scientific Reports.  2017 Sep; 7(1).
\newblock \urlprefix\url{http://dx.doi.org/10.1038/s41598-017-10421-y},
  \href{10.1038/s41598-017-10421-y}{\doiprefix
  \detokenize{10.1038/s41598-017-10421-y}}.

\bibitem[{Kikuchi et~al.(1991)K. Kikuchi and M. Yoshida and T. Maekawa and H.
  Watanabe}]{kikuchi1991}
\textbf{\color{eLifeMediumGrey} Kikuchi K}, Yoshida M, Maekawa T, Watanabe H.
\newblock Metropolis Monte Carlo method as a numerical technique to solve the
  Fokker{\textemdash}Planck equation.
\newblock Chemical Physics Letters.  1991 Oct; 185(3-4):335--338.
\newblock \urlprefix\url{https://doi.org/10.1016/s0009-2614(91)85070-d},
  \href{10.1016/s0009-2614(91)85070-d}{\doiprefix
  \detokenize{10.1016/s0009-2614(91)85070-d}}.

\bibitem[{Kilic et~al.(2021)Kilic, Zeliha and Sgouralis, Ioannis and Pressé,
  Steve}]{rnap_residence}
\textbf{\color{eLifeMediumGrey} Kilic Z}, Sgouralis I, Pressé S.
\newblock Residence time analysis of RNA polymerase transcription dynamics: A
  Bayesian sticky HMM approach.
\newblock Biophysical Journal.  2021 May; 120(9):1665–1679.
\newblock \urlprefix\url{http://dx.doi.org/10.1016/j.bpj.2021.02.045},
  \href{10.1016/j.bpj.2021.02.045}{\doiprefix
  \detokenize{10.1016/j.bpj.2021.02.045}}.

\bibitem[{Ladouceur et~al.(2020)Ladouceur, Anne-Marie and Parmar, Baljyot Singh
  and Biedzinski, Stefan and Wall, James and Tope, S. Graydon and Cohn, David
  and Kim, Albright and Soubry, Nicolas and Reyes-Lamothe, Rodrigo and Weber,
  Stephanie C.}]{rnap_llps}
\textbf{\color{eLifeMediumGrey} Ladouceur AM}, Parmar BS, Biedzinski S, Wall J,
  Tope SG, Cohn D, Kim A, Soubry N, Reyes-Lamothe R, Weber SC.
\newblock Clusters of bacterial {RNA} polymerase are biomolecular condensates
  that assemble through liquid–liquid phase separation.
\newblock Proceedings of the National Academy of Sciences.  2020 Jul;
  117(31):18540–18549.
\newblock \urlprefix\url{http://dx.doi.org/10.1073/pnas.2005019117},
  \href{10.1073/pnas.2005019117}{\doiprefix
  \detokenize{10.1073/pnas.2005019117}}.

\bibitem[{Lafontaine et~al.(2020)Lafontaine, Denis L. J. and Riback, Joshua A.
  and Bascetin, R\"{u}meyza and Brangwynne, Clifford P.}]{nucleolus3}
\textbf{\color{eLifeMediumGrey} Lafontaine DLJ}, Riback JA, Bascetin R,
  Brangwynne CP.
\newblock The nucleolus as a multiphase liquid condensate.
\newblock Nature Reviews Molecular Cell Biology.  2020 Sep; 22(3):165–182.
\newblock \urlprefix\url{http://dx.doi.org/10.1038/s41580-020-0272-6},
  \href{10.1038/s41580-020-0272-6}{\doiprefix
  \detokenize{10.1038/s41580-020-0272-6}}.

\bibitem[{Lieberman-Aiden et~al.(2009)Lieberman-Aiden, Erez and van Berkum,
  Nynke L. and Williams, Louise and Imakaev, Maxim and Ragoczy, Tobias and
  Telling, Agnes and Amit, Ido and Lajoie, Bryan R. and Sabo, Peter J. and
  Dorschner, Michael O. and Sandstrom, Richard and Bernstein, Bradley and
  Bender, M. A. and Groudine, Mark and Gnirke, Andreas and Stamatoyannopoulos,
  John and Mirny, Leonid A. and Lander, Eric S. and Dekker, Job}]{hic_original}
\textbf{\color{eLifeMediumGrey} Lieberman-Aiden E}, van Berkum NL, Williams L,
  Imakaev M, Ragoczy T, Telling A, Amit I, Lajoie BR, Sabo PJ, Dorschner MO,
  Sandstrom R, Bernstein B, Bender MA, Groudine M, Gnirke A, Stamatoyannopoulos
  J, Mirny LA, Lander ES, Dekker J.
\newblock Comprehensive Mapping of Long-Range Interactions Reveals Folding
  Principles of the Human Genome.
\newblock Science.  2009 Oct; 326(5950):289–293.
\newblock \urlprefix\url{http://dx.doi.org/10.1126/science.1181369},
  \href{10.1126/science.1181369}{\doiprefix
  \detokenize{10.1126/science.1181369}}.

\bibitem[{Lioy et~al.(2018)Virginia S. Lioy and Axel Cournac and Martial
  Marbouty and St{\'{e}}phane Duigou and Julien Mozziconacci and Olivier
  Esp{\'{e}}li and Fr{\'{e}}d{\'{e}}ric Boccard and Romain Koszul}]{Lioy2018}
\textbf{\color{eLifeMediumGrey} Lioy VS}, Cournac A, Marbouty M, Duigou S,
  Mozziconacci J, Esp{\'{e}}li O, Boccard F, Koszul R.
\newblock Multiscale Structuring of the {E.~coli} Chromosome by
  Nucleoid-Associated and Condensin Proteins.
\newblock Cell.  2018 Feb; 172(4):771--783.e18.
\newblock \urlprefix\url{https://doi.org/10.1016/j.cell.2017.12.027},
  \href{10.1016/j.cell.2017.12.027}{\doiprefix
  \detokenize{10.1016/j.cell.2017.12.027}}.

\bibitem[{Mangiameli et~al.(2018)Sarah M. Mangiameli and Julie A. Cass and
  Houra Merrikh and Paul A. Wiggins}]{Wiggins2018}
\textbf{\color{eLifeMediumGrey} Mangiameli SM}, Cass JA, Merrikh H, Wiggins PA.
\newblock The bacterial replisome has factory-like localization.
\newblock Current Genetics.  2018 Apr; 64(5):1029--1036.
\newblock \urlprefix\url{https://doi.org/10.1007/s00294-018-0830-z},
  \href{10.1007/s00294-018-0830-z}{\doiprefix
  \detokenize{10.1007/s00294-018-0830-z}}.

\bibitem[{Mukherjee et~al.(2026)Mukherjee, Sukanta and Skoruppa, Enrico and
  Merlitz, Holger and Sommer, Jens-Uwe and Schiessel, Helmut}]{Mukherjee2026}
\textbf{\color{eLifeMediumGrey} Mukherjee S}, Skoruppa E, Merlitz H, Sommer JU,
  Schiessel H.
\newblock A Self-Organized Liquid Reaction Container for Cellular Memory.
\newblock Advanced Science.  2026; 13(20):e12500.
\newblock
  \urlprefix\url{https://advanced.onlinelibrary.wiley.com/doi/abs/10.1002/advs.202512500},
  \href{https://doi.org/10.1002/advs.202512500}{\doiprefix
  \detokenize{https://doi.org/10.1002/advs.202512500}}.

\bibitem[{Pelletier et~al.(2012)Pelletier, James and Halvorsen, Ken and Ha,
  Bae-Yeun and Paparcone, Raffaella and Sandler, Steven J. and Woldringh,
  Conrad L. and Wong, Wesley P. and Jun, Suckjoon}]{Pelletier2012}
\textbf{\color{eLifeMediumGrey} Pelletier J}, Halvorsen K, Ha BY, Paparcone R,
  Sandler SJ, Woldringh CL, Wong WP, Jun S.
\newblock Physical manipulation of the {Escherichia} coli chromosome reveals
  its soft nature.
\newblock Proceedings of the National Academy of Sciences.  2012 Sep; 109(40).
\newblock \urlprefix\url{http://dx.doi.org/10.1073/pnas.1208689109},
  \href{10.1073/pnas.1208689109}{\doiprefix
  \detokenize{10.1073/pnas.1208689109}}.

\bibitem[{Phillips et~al.(2012)Phillips, Rob and Kondev, Jane and Theriot,
  Julie and Garcia, Hernan G. and Orme, Nigel}]{Phillips2012}
\textbf{\color{eLifeMediumGrey} Phillips R}, Kondev J, Theriot J, Garcia HG,
  Orme N.
\newblock Physical Biology of the Cell.
\newblock Garland Science; 2012.
\newblock \urlprefix\url{http://dx.doi.org/10.1201/9781134111589},
  \href{10.1201/9781134111589}{\doiprefix \detokenize{10.1201/9781134111589}}.

\bibitem[{Polovnikov and Starkov(2026)Polovnikov, Kirill and Starkov,
  Dmitry}]{Polovnikov2026}
\textbf{\color{eLifeMediumGrey} Polovnikov K}, Starkov D.
\newblock A universal polymer signature in {Hi-C} resolves cohesin loop density
  and supports monomeric extrusion.
\newblock Proceedings of the National Academy of Sciences.  2026 Apr; 123(16).
\newblock \urlprefix\url{http://dx.doi.org/10.1073/pnas.2534385123},
  \href{10.1073/pnas.2534385123}{\doiprefix
  \detokenize{10.1073/pnas.2534385123}}.

\bibitem[{Postow et~al.(2004)Postow, Lisa and Hardy, Christine D. and Arsuaga,
  Javier and Cozzarelli, Nicholas R.}]{supercoiled_loop}
\textbf{\color{eLifeMediumGrey} Postow L}, Hardy CD, Arsuaga J, Cozzarelli NR.
\newblock Topological domain structure of the {Escherichia} coli chromosome.
\newblock Genes  and  Development.  2004 Jul; 18(14):1766–1779.
\newblock \urlprefix\url{http://dx.doi.org/10.1101/gad.1207504},
  \href{10.1101/gad.1207504}{\doiprefix \detokenize{10.1101/gad.1207504}}.

\bibitem[{Ripin and Parker(2023)Ripin, Nina and Parker, Roy}]{Ripin2023}
\textbf{\color{eLifeMediumGrey} Ripin N}, Parker R.
\newblock Formation, function, and pathology of {RNP} granules.
\newblock Cell.  2023 Oct; 186(22):4737–4756.
\newblock \urlprefix\url{http://dx.doi.org/10.1016/j.cell.2023.09.006},
  \href{10.1016/j.cell.2023.09.006}{\doiprefix
  \detokenize{10.1016/j.cell.2023.09.006}}.

\bibitem[{Romano et~al.(2011)Flavio Romano and Cristiano De Michele and Davide
  Marenduzzo and Eduardo Sanz}]{Romano2011}
\textbf{\color{eLifeMediumGrey} Romano F}, Michele CD, Marenduzzo D, Sanz E.
\newblock Monte Carlo and event-driven dynamics of Brownian particles with
  orientational degrees of freedom.
\newblock The Journal of Chemical Physics.  2011 Sep; 135(12):124106.
\newblock \urlprefix\url{https://doi.org/10.1063/1.3629452},
  \href{10.1063/1.3629452}{\doiprefix \detokenize{10.1063/1.3629452}}.

\bibitem[{Shin and Brangwynne(2017)Shin, Yongdae and Brangwynne, Clifford
  P.}]{Shin2017}
\textbf{\color{eLifeMediumGrey} Shin Y}, Brangwynne CP.
\newblock Liquid phase condensation in cell physiology and disease.
\newblock Science.  2017 Sep; 357(6357).
\newblock \urlprefix\url{http://dx.doi.org/10.1126/science.aaf4382},
  \href{10.1126/science.aaf4382}{\doiprefix
  \detokenize{10.1126/science.aaf4382}}.

\bibitem[{Singh et~al.(2017)Awaneesh Singh and Amrita Singh and Anirban
  Chakraborti}]{Singh2017}
\textbf{\color{eLifeMediumGrey} Singh A}, Singh A, Chakraborti A.
\newblock Effect of bond-disorder on the phase-separation kinetics of binary
  mixtures: {A Monte Carlo} simulation study.
\newblock The Journal of Chemical Physics.  2017 Sep; 147(12):124902.
\newblock \urlprefix\url{https://doi.org/10.1063/1.5004563},
  \href{10.1063/1.5004563}{\doiprefix \detokenize{10.1063/1.5004563}}.

\bibitem[{Sommer et~al.(2022)Sommer, Jens-Uwe and Merlitz, Holger and
  Schiessel, Helmut}]{Sommer_PAC}
\textbf{\color{eLifeMediumGrey} Sommer JU}, Merlitz H, Schiessel H.
\newblock Polymer-Assisted Condensation: A Mechanism for Hetero-Chromatin
  Formation and Epigenetic Memory.
\newblock Macromolecules.  2022 May; 55(11):4841–4851.
\newblock \urlprefix\url{http://dx.doi.org/10.1021/acs.macromol.2c00244},
  \href{10.1021/acs.macromol.2c00244}{\doiprefix
  \detokenize{10.1021/acs.macromol.2c00244}}.

\bibitem[{Strom et~al.(2017)Strom, Amy R. and Emelyanov, Alexander V. and Mir,
  Mustafa and Fyodorov, Dmitry V. and Darzacq, Xavier and Karpen, Gary
  H.}]{Strom2017}
\textbf{\color{eLifeMediumGrey} Strom AR}, Emelyanov AV, Mir M, Fyodorov DV,
  Darzacq X, Karpen GH.
\newblock Phase separation drives heterochromatin domain formation.
\newblock Nature.  2017 Jun; 547(7662):241–245.
\newblock \urlprefix\url{http://dx.doi.org/10.1038/nature22989},
  \href{10.1038/nature22989}{\doiprefix \detokenize{10.1038/nature22989}}.

\bibitem[{Trueba and Woldringh(1980)Trueba, F J and Woldringh, C L}]{Trueba869}
\textbf{\color{eLifeMediumGrey} Trueba FJ}, Woldringh CL.
\newblock Changes in cell diameter during the division cycle of Escherichia
  coli.
\newblock Journal of Bacteriology.  1980; 142(3):869--878.
\newblock \urlprefix\url{https://jb.asm.org/content/142/3/869}.

\bibitem[{Viollier et~al.(2004)P. H. Viollier and M. Thanbichler and P. T.
  McGrath and L. West and M. Meewan and H. H. McAdams and L.
  Shapiro}]{caul_loci}
\textbf{\color{eLifeMediumGrey} Viollier PH}, Thanbichler M, McGrath PT, West
  L, Meewan M, McAdams HH, Shapiro L.
\newblock Rapid and sequential movement of individual chromosomal loci to
  specific subcellular locations during bacterial {DNA} replication.
\newblock Proceedings of the National Academy of Sciences.  2004 Jun;
  101(25):9257--9262.
\newblock \urlprefix\url{https://doi.org/10.1073/pnas.0402606101},
  \href{10.1073/pnas.0402606101}{\doiprefix
  \detokenize{10.1073/pnas.0402606101}}.

\bibitem[{Weber et~al.(2010)Stephanie C. Weber and Andrew J. Spakowitz and
  Julie A. Theriot}]{Weber2010}
\textbf{\color{eLifeMediumGrey} Weber SC}, Spakowitz AJ, Theriot JA.
\newblock Bacterial Chromosomal Loci Move Subdiffusively through a Viscoelastic
  Cytoplasm.
\newblock Physical Review Letters.  2010 Jun; 104(23).
\newblock \urlprefix\url{https://doi.org/10.1103/physrevlett.104.238102},
  \href{10.1103/physrevlett.104.238102}{\doiprefix
  \detokenize{10.1103/physrevlett.104.238102}}.

\bibitem[{Weng et~al.(2019)Weng, Xiaoli and Bohrer, Christopher H. and
  Bettridge, Kelsey and Lagda, Arvin Cesar and Cagliero, Cedric and Jin, Ding
  Jun and Xiao, Jie}]{xiao19}
\textbf{\color{eLifeMediumGrey} Weng X}, Bohrer CH, Bettridge K, Lagda AC,
  Cagliero C, Jin DJ, Xiao J.
\newblock Spatial organization of RNA polymerase and its relationship with
  transcription in {Escherichia} coli.
\newblock Proceedings of the National Academy of Sciences.  2019 Sep;
  116(40):20115–20123.
\newblock \urlprefix\url{http://dx.doi.org/10.1073/pnas.1903968116},
  \href{10.1073/pnas.1903968116}{\doiprefix
  \detokenize{10.1073/pnas.1903968116}}.

\bibitem[{Wiggins et~al.(2010)Wiggins, Paul A. and Cheveralls, Keith C. and
  Martin, Joshua S. and Lintner, Robert and Kondev, Jané}]{Wiggins2010}
\textbf{\color{eLifeMediumGrey} Wiggins PA}, Cheveralls KC, Martin JS, Lintner
  R, Kondev J.
\newblock Strong intranucleoid interactions organize the {Escherichia} coli
  chromosome into a nucleoid filament.
\newblock Proceedings of the National Academy of Sciences.  2010 Mar;
  107(11):4991–4995.
\newblock \urlprefix\url{http://dx.doi.org/10.1073/pnas.0912062107},
  \href{10.1073/pnas.0912062107}{\doiprefix
  \detokenize{10.1073/pnas.0912062107}}.

\bibitem[{Yoneda et~al.(2021)Yoneda, Mitsuhiro and Nakagawa, Takeya and
  Hattori, Naoko and Ito, Takashi}]{nucleolus}
\textbf{\color{eLifeMediumGrey} Yoneda M}, Nakagawa T, Hattori N, Ito T.
\newblock The nucleolus from a liquid droplet perspective.
\newblock The Journal of Biochemistry.  2021 08; 170(2):153--162.
\newblock \urlprefix\url{https://doi.org/10.1093/jb/mvab090},
  \href{10.1093/jb/mvab090}{\doiprefix \detokenize{10.1093/jb/mvab090}}.

\bibitem[{Youngren et~al.(2014)B. Youngren and H. J. Nielsen and S. Jun and S.
  Austin}]{Youngren2014}
\textbf{\color{eLifeMediumGrey} Youngren B}, Nielsen HJ, Jun S, Austin S.
\newblock The multifork {Escherichia} coli chromosome is a self-duplicating and
  self-segregating thermodynamic ring polymer.
\newblock Genes and Development.  2014 Jan; 28(1):71--84.
\newblock \urlprefix\url{https://doi.org/10.1101/gad.231050.113},
  \href{10.1101/gad.231050.113}{\doiprefix
  \detokenize{10.1101/gad.231050.113}}.

\end{thebibliography}

\end{document}